\documentclass[oldversion]{aa}
\usepackage{txfonts}
\usepackage{natbib}
 \usepackage{times}
 \usepackage{amsfonts}
 \usepackage{amssymb}
 \usepackage{graphicx}

\newcommand{\sub}[1]{_{\rm #1}}

\begin{document}

\title{[CII] 158$\mu$m Emission and Metallicity in PDRs}
\titlerunning{[CII] 158$\mu$m Emission and Metallicity in PDRs}
\author{M. R\"ollig\inst{1},V. Ossenkopf\inst{1,2}, S. Jeyakumar,  J. Stutzki\inst{1}, and A.~Sternberg\inst{3} }
\authorrunning{R\"ollig et al. }
\institute{I. Physikalisches Institut, Universit\"at zu K\"oln, Z\"ulpicher Str. 77, D-50937 K\"oln,
Germany
\and SRON National Institute for Space Research, P.O. Box 800, 9700 AV Groningen, the Netherlands
\and School of Physics and Astronomy, Tel Aviv University, Ramat Aviv 69978, Israel}

\offprints{M.R\"ollig,\\
 \email{roellig@ph1.uni-koeln.de}}

\abstract{
We study the effects of a metallicity variation on
the thermal balance and [CII] fine-structure line
strengths in
interstellar photon dominated regions (PDRs).
We find that a reduction in the dust-to-gas ratio and
the abundance of heavy elements in the gas phase
changes the heat balance of the gas in PDRs. 
The surface temperature of PDRs decreases as
the metallicity decreases except for high density
($n>10^6$ cm$^{-3}$) clouds exposed to weak ($\chi< 100$) FUV fields where
vibrational
H$_2$-deexcitation heating dominates over photoelectric heating of the gas.
We incorporate the metallicity dependence in our KOSMA-$\tau$ PDR model 
to study the metallicity dependence of [CII]/CO line ratios
in low metallicity galaxies.  We find that the main trend in the variation
of the observed CII/CO ratio with metallicity is well reproduced
by a single spherical clump, and does not necessarily require
an ensemble of clumps as in the semi-analytical model presented
by Bolatto et al. (1999).

}

\maketitle

\section{Introduction}
The bright line emission from photon-dominated regions (PDRs) is one of the
key tracers of the star formation activity throughout the evolution of 
galaxies in the course of the
cosmological evolution. Hence, proper modeling of PDR emission is of central
importance for the interpretation of the observations, in order to derive 
the physical parameters and the chemical state
of the ISM in external galaxies. The extensive literature on PDR emission,
both observationally and from the modeling side, has largely concentrated on
bright Galactic sources and starburst galaxies. The effect of different 
metallicity for the resulting PDR emission has, up to now, drawn little
attention. It is, however, very important in order to cover the full course of
galactic evolution, starting with low metallicity material of cosmological origin.
Many nearby galaxies, such as dwarf galaxies, irregular galaxies and the Magellanic
Clouds have a low metallicity \citep{lisenfeld98, kunth00, pustilnik02, lee03}.
Within the Galaxy, as well as in other spiral galaxies,
there is a radial decrease in the metallicity
of molecular clouds and associated H{\small II} regions 
\citep{zaritsky94, arimoto96, giveon, bresolin04}. 
These systems provide the
opportunity to study star formation and photon-dominated regions (PDRs)
for a variety of metallicities.

In PDRs the molecular gas
is heated by the far-ultraviolet (FUV) radiation field, either 
the strong FUV radiation in the vicinity of hot young stars, or weak average
FUV fields in the Galaxy. 
The gas cools through the spectral line radiation of atomic and
molecular species (Hollenbach \& Tielens 1999, Sternberg 2004).
The gas-phase chemistry together with a depth dependent FUV intensity
lead to the formation of atomic and molecular species at different depths through
the cloud. This typical stratification of PDRs is for example reflected by the the 
H/H$_2$ and C$^+$/C/CO transitions \citep{SD95, boger05}.
At low visual extinctions the gas
is cooled by emission of atomic fine-structure lines, mainly [CII] 158$\mu$m and
[OI] 63$\mu$m. At larger depths,
millimeter, sub-millimeter and far-infrared
molecular rotational-line cooling (CO, OH, H$_2$O) becomes important together 
with the interaction of dust and gas. 
Physical conditions such as temperature and density
can be derived, by  comparing the observed line emissions with model predictions
 \citep{lebourlot93, wolfire95, stoerzer96, warin96, kaufman99, zielinsky00, stoerzer00, gorti02}.

[CII] emission is a widely used diagnostic
indicator of star formation \citep{stacey91, pierini99, malhotra00, boselli02b, pierini03, kramer04}.
Observations suggest that low metallicity systems have
higher [CII] to CO rotational line ratios
compared to the Galactic value. In particular, the intensity ratio
$I_{\mathrm{[CII]}}/I_{\mathrm{CO}}$ may vary from $\sim 1000$ in
 the inner Milky Way, up to $\sim 10^5$ in extremely low metallicity systems
\citep[eg.][]{madden97, mochi98, bolatto99, madden00, hunter01}.
Several studies have suggested that a lower abundance 
of heavy elements affects the chemical structure
of PDRs and the cooling line emission, and
that estimates of molecular gas masses from the
observed CO(J=1-0) line intensities using the standard conversion factor may
underestimate the true masses
in such objects \citep{wilson95, israel97, israel03, rubio04}.

\citet{bolatto99} modelled the metallicity variation of the line ratio [CII]/CO(1-0),
for an ensemble of spherical ``clumps'', assuming an
inverse relation between the size of the C$^+$ region
and the metallicity.
However the sizes of the C$^+$, C and CO regions also depend on the chemistry in PDRs
and the chemical network is modified at low metallicities \citep{lequex94}. Additionally it 
has been suggested that the size of the C$^+$, C and CO regions may also  significantly
depend on the overall cloud morphology, e.g. density variations \citep{hegmann03} and velocity
 fluctuations \citep{roellig02}.    
Moreover the temperature of the molecular gas might depend on the metallicity
which affects the observable line intensities \citep{wolfire95}.

We study the effects of metallicity changes on the temperature and chemical structure
of PDRs.
In \S~\ref{metalsection} we consider the dependence of the PDR gas
temperature on the metallicity using a simplified semi-analytic model
and compare it with numerical results from full 
PDR model calculations. Our computations were carried out
using an updated version of our spherical 
KOSMA-$\tau$ model \citep{stoerzer96} which was originally adapted from
the plane-parallel model presented by Sternberg \& Dalgarno (1995).  In \S~\ref{ciiw} we examine the
predicted size of the C$^+$ zones as a function of metallicity.
We then model the strength of the [CII] emission and investigate the
the dependence of the [CII]/CO(J=1-0) line ratio on the metallicity.
Finally we compare the results with observational data in \S~\ref{clumpysection}.

\section{Metallicity dependence of the surface temperature}
\label{metalsection}

The basic  cooling and heating processes in PDRs, are affected
by the abundances of elements as well as the content and the composition of dust grains
\citep{wolfire95, kaufman99}.
The dust-to-gas ratio ($D/G$) and the optical properties of the dust
may depend on the metallicity, Z. Fits to observations suggest that 
the ratio depends almost linearly on the metallicity, $D/G\,\propto Z^{1.146}$
\citep{boselli02a}.  There are other studies that find deviations from linearity
for higher values of  $D/G$ \citep{lisenfeld98}. \citet{li02} suggested that 
the mixture of PAHs in the metal-poor SMC differs from the Milky Way. 
There are a few observations indicating that PAHs could have
been destroyed by intense UV fields at low metallicities \citep{thuan99, bolatto00},
but the detailed composition of dust in low metallicity
environments and the influence on its optical properties is not yet understood.
Because of the insufficient knowledge we assume in our model that
the composition of the grains does not change with
metallicity and that the dust-to-gas ratio and the gas-phase abundance
of heavy elements scale linearly with $Z$.

Changes in $Z$ affect the
abundances of major coolants as well as the electron densities 
in PDRs. Additionally, a reduction
in the dust abundance diminishes the  UV opacity,
the photo electric heating rate, and the H$_2$ formation rate.
These changes affect the temperature and chemistry in the surface layers
where C$^+$ is most abundant.

The dependence of the surface gas temperature on $Z$ can be
estimated considering the balance of cooling and heating.
The dominant cooling processes depend predominantly on the total hydrogen
gas density $n$.
[OI]63$\mu m$, [CII]158$\mu m$ emission, and gas-grain collisions 
are important cooling processes \citep{burke83, stoerzer96}. Their relative importance
in the different regimes is discussed in Sect. \ref{sect_cooling}.
The dominant heating process depends on the far-ultraviolet (FUV; 6-13.6 eV) 
field and the density. Grain photo-electric emission
(PE) \citep{dhend87, lepp88, ver90, BT94}, collision deexcitation
of FUV pumped molecular hydrogen H$_2^\star$ \citep{SD89, burton90} and 
heating from H$_2$ formation play important roles. They are discussed in
detail in Sect. \ref{sect_heating}. By explicitly considering the
metallicity dependence of each of these cooling and heating
processes and identifying the dominant processes in the different
parameter regimes we will show how the energy balance
in PDRs depends on $Z$ for a quantitative
understanding of the PDR surface temperature.

\subsection{The KOSMA-$\tau$ PDR model}
\label{pdrmodel}
In our study 
we use an updated version of the spherical PDR code described
in detail by \citet{stoerzer96}.
Briefly, this model solves the coupled equations of energy balance
(heating and cooling), chemical equilibrium, and radiative transfer 
in spherical geometry.
The PDR-clumps are characterized by 
a) the incident FUV field 
intensity $\chi$, given in units of the mean interstellar radiation
field of Draine (1978),
b) the clump mass, and c) the average density of the clump, for a
radial power-law density distribution with index $\gamma$.
We incorporate the effects of varying metallicity by varying the
assumed abundance of dust grains and
heavy elements.  The following parameters are multiplied by the
metallicity factor $Z$: (a) the total effective FUV dust absorption cross section
per hydrogen nucleus $\sigma$;
(b) the photo-electric heating rate;
(c) the H$_2$ formation rate;
(d) the metal abundance. We consider a range of $Z$ from 0.2 to 1.
For $Z=1$ we use $\sigma=1.9\times 10^{-21}$~cm$^2$,
C/H=$1.4\times 10^{-4}$, and O/H=$3\times 10^{-4}$ as standard values for the 
local ISM \citep{HT99}. These values are slightly lower than recent
solar values of $12+\log (\mathrm{O/H})\approx8.7$ by \cite{asplund04}. For a 
detailed discussion see \cite{baumgartner05}.  We do not 
consider gas depletion on grains.

\subsection{Semi-analytic approximations}
The results of the full numerical computations can be understood and anticipated
using some simplifying semi-analytical approximations. 
Here, we focus on the surface temperature of the PDR at $A_V$=0, 
as we are not interested in the shielding properties of low-metal
PDRs but in the thermal behavior of the low-extinction region dominating the
CII emission. We study how the thermal properties respond to altered elemental abundance.
We assume the molecular cloud is sufficiently thick such that it
absorbs all radiation coming from the backside, leading to
an emission line escape probability $\beta(\tau=0)=1/2$ at the surface. 

\subsubsection{Cooling}
\label{sect_cooling}
In the simplified model we
include the three main cooling processes: [CII] and [OI] line cooling 
and gas-grain collisional cooling \citep{burke83, stoerzer96}. 
Gas cooling is generally 
dominated by fine structure emission of [CII] and [OI]. Gas-grain 
cooling starts to contribute significantly for high densities. 
At $n\gtrsim10^6$~cm$^{-3}$ the coupling between gas and dust is strong
enough so that the efficient cooling of the dust by infrared radiation 
also provides a major coolant to the gas.
The total cooling rate per unit volume by these radiative processes  
is the sum
\begin{equation}
\Lambda_\mathrm{tot}=\Lambda_\mathrm{CII}+\Lambda_\mathrm{
OI}+\Lambda_\mathrm{g-g}
\end{equation}
With the abundance of carbon, oxygen, and dust scaling with the metallicity,
the total cooling rate is also linear in $Z$.

Analytic expressions for the three cooling processes are 
derived 
in Appendix A.  In Figure (\ref{bal_cool}) we show the
relation between the fine-structure line cooling contributions
as a function of $n$ and $\chi$ for two different values of $Z$. 
The region in parameter space where [CII] cooling
dominates over [OI] is shaded in gray. For densities above $~10^{4.5}$ cm$^{-3}$
[CII] cooling is quenched and [OI] cooling 
dominates. At all lower densities [CII] cooling dominates.
Every point in Figure \ref{bal_cool} corresponds to a different
equilibrium temperature and resulting ratio 
$\Lambda_\mathrm{CII}/\Lambda_\mathrm{OI}$.
The dependence on $\chi$ in Figure \ref{bal_cool} results from 
the implicit temperature dependence of the fine-structure cooling rates.
 Grain cooling  dominates
only for densities greater than $10^6$~cm$^{-3}$.

 \begin{figure}[htp]
\centering
\resizebox{\hsize}{!}{\includegraphics[width=14cm]{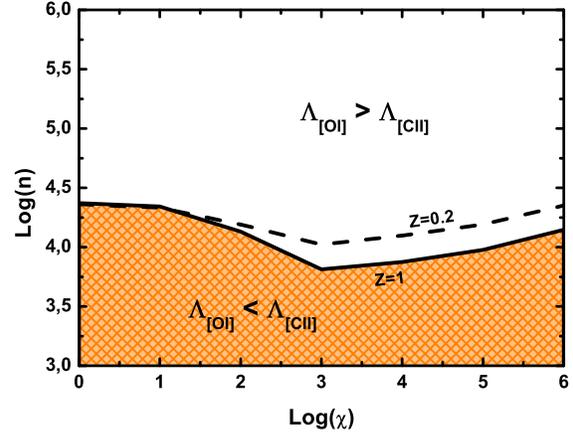}}
 \caption{The solid and dashed line represent the points in $n-\chi$-parameter space where $\Lambda_\mathrm{OI}=\Lambda_\mathrm{CII}$ for 
metallicities of $Z=1$ and $Z=0.2$ respectively.  \label{bal_cool} }
\end{figure}
\begin{figure}[htp]
\centering
\resizebox{\hsize}{!}{\includegraphics[width=14cm]{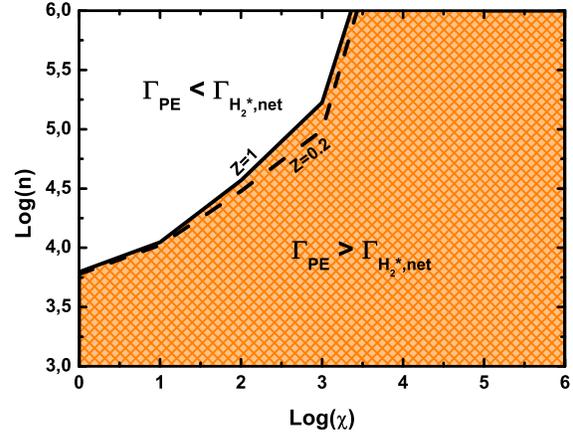}}
 \caption{The solid and dashed line represent the points in $n-\chi$-parameter space where $\Gamma_\mathrm{H_2^*,net}=\Gamma_\mathrm{PE}$ for 
metallicities of $Z=1$ and $Z=0.2$ respectively.  \label{bal_heat} }

 \end{figure}

\subsubsection{Heating}
\label{sect_heating}
The dominant heating process depends on the FUV
field intensity and density. For high intensities grain photo-electric heating
dominates. The rate for this process, given by \citet{BT94}, is
$\Gamma_{\mathrm PE}=10^{-24} \epsilon \,G_0n_{\mathrm H}\,Z$
erg~s$^{-1}$~cm$^{-3}$ where $\epsilon$ is the photoelectric
heating efficiency and $G_0$ is the UV intensity in units of the
Habing field. Following \citet{BT94} the photoelectric heating
efficiency is given by:
\begin{equation}
\epsilon = \frac{3\times 10^{-2}}{1+2\times10^{-4}(G_0
T^{1/2}/n_e)}
\end{equation}
with $n_e$ being the electron density in cm$^{-3}$ and $T$ the
dust temperature in K. We set $G_0=1.71\times 0.5\,\chi$ to account for the relative
factor of 1.71 between the Habing and Draine fields, and the fact that
at the surface of optically thick clouds radiation is incident
from a solid angle of $2\,\pi$ rather than $4\,\pi$ steradians.
In evaluating the efficiency $\epsilon$ we use the
analytic expression for the electron density derived in Appendix \ref{appelectron}:
\begin{equation}\label{densityelectron}
n_e\approx 0.84\times 10^{-4}\, n\, Z\,
\left(1+\sqrt{1+14.4\,\frac{T^{0.75}}{n\, Z^2}}\,\right)  \;{\rm
cm^{-3}}
\end{equation}

It is common to express the {\it net} PE heating rate as 
$\Gamma_\mathrm{PE}^{net}=\Gamma_\mathrm{PE}-\Lambda_\mathrm{rec}$, where 
$\Lambda_\mathrm{rec}$ is the cooling rate due to electron recombination. 
We adopt the analytical fit from  \citet{BT94}:
\begin{equation}
 \Lambda_{rec}=3.49\times 10^{-30}\,T^{0.944}\,\left(\frac{G_0\,T^{1/2}}{n_e}\right)^{\frac{0.735}{T^{0.068}}}n_e\, n\, Z
\end{equation}

In dense PDRs a second important heating source is the collisional
deexcitation of 
vibrationally excited H$_2^\star$. The rate for this process 
can be expressed as
\begin{equation}\label{h2hone}
\Gamma_{\mathrm H_2^\star}=\chi\,P\,n_{\mathrm
 H_2}\, \Delta E\, f \;\;\;\;{\rm erg\, s^{-1} cm^{-3}},
\end{equation}
with $n_{\mathrm H_2}$ is the density of molecular hydrogen , 
$P=2.9\times 10^{-10}$~s$^{-1}$ is the pumping rate for a unit FUV field, $\Delta E\approx 23500$~K is the characteristic
vibrational transition energy, and an
efficiency factor $f$ accounting for all processes that may reduce
the number of de-exciting collisions (see Appendix B).
The balance equation for the formation and destruction of H$_2$ is
\begin{equation}\label{nh1}
n\,n_{\mathrm H}\,R=\chi\,D\,n_{\mathrm H_2}
\end{equation}    
where $D=2.6\times 10^{-11}$~s$^{-1}$ is the total dissociation rate in a $\chi=1$ FUV field, and 
$R=R_0\, Z$ is the grain surface H$_2$ formation rate coefficient (cm$^3$~s$^{-1}$).
Here, it is implicitly assumed that dust grains are always
covered by enough $H$ atoms, so that the recombination rate
is only limited by the number of H-dust collisions.
We use the standard recombination rate $R_0$ by \citet{HS71a,HS71b}:
\begin{equation}\label{rform}
R_0=3\times10^{-18}\,f_a\,S\,T^{1/2} \; {\rm cm^{3}\,s^{-1}}.
\end{equation}
where
the accommodation coefficient, $f_a$ and the sticking probability
$S$ are independent of $Z$ \citep{HM79}. \citet{cazaux04b} show
that gas-phase formation of H$_2$ becomes important for $Z<10^{-3}$, well below
 the minimum value of $Z$ we consider here.

The density of atomic and molecular hydrogen
is determined by Eq.(\ref{nh1}) and can be written as
\begin{equation}\label{alpha}
n_\mathrm{H}=n\frac{1}{1+2\,\alpha}\,\,\, \mbox{and}\,\,\, n_\mathrm{H_2}=n\frac{\alpha}{1+2\,\alpha}
\end{equation}
respectively, with $\alpha=n\,R/(\chi\,D)$ being the
 ratio between formation and destruction rate coefficients. 
For example, for $\chi=1$ and $n=10^3$~cm$^{-3}$ more than 99\% of the 
gas at the surface is atomic. 
For $Z=1$ and a unit Draine field, $\alpha=1$ and $n_\mathrm{H}=n_\mathrm{H_2}$,  for
densities $n\approx10^{6}$~cm$^{-3}$. As only  
$R$ depends on $Z$, it follows that $\alpha \propto Z$ and the 
density $n$ at which the H and H$_2$ densities are equal at 
the cloud surface scales as $1/Z$.

From
equations \ref{h2hone}, \ref{nh1}, and \ref{rform} it follows that
\begin{equation}\label{h2htwo}
\Gamma_{\mathrm H_2^\star}=\left(\frac{P}{D}\right)\, R_0\,Z\, n\, n_{\mathrm H}\, \Delta E\, f\;\;\; {\rm erg\, s^{-1} cm^{-3}}.
\end{equation}
We see that 
$\Gamma_{\mathrm H_2^\star}$ is maximized when $n_\mathrm{H}=n$ and all of 
the hydrogen is atomic. The efficiency factor $f$ in Equation~\ref{h2hone}
is largest in the limit of high gas density and low FUV intensity, where
radiative processes become negligible in depopulating the excited vibrational
levels compared to collisional deexcitation. For a temperature 
of $T=100$~K, $Z=1$, $\chi=1$, and $n=10^3$~cm$^{-3}$ we obtain
$\Gamma_{\mathrm H_2^\star}^{max}\approx 1.8\times 10^{-24} \; {\rm erg\,cm^{-3}\,s^{-1}}$
with $f=2.8\times 10^{-3}$.

This is in good agreement with the results from the numerical PDR
model shown in Fig.~\ref{h2hn30} in Appendix \ref{apph2heating}.
The assumption of a constant formation rate $R_0$ is valid for 
$\chi\lesssim 10^3$. A higher UV field the dust temperature increases
leading to a rapid reduction of the accommodation coefficient $f_a$. 
Hence in our calculations the maximum H$_2$ heating rate drops for $\chi\gtrsim 10^3$
as shown in the bottom plot in Fig.~\ref{h2hn30} in Appendix \ref{apph2heating}.

H$_2$ not only contributes to the heating, but  
cools the gas at higher temperatures \citep{SD89}. To account for the cooling we 
define the net heating rate $\Gamma_\mathrm{H_2}^{net}=\Gamma_\mathrm
{H_2^\star}-\Lambda_\mathrm{H_2}$. Using the analytic approximations
to the molecular level structure of H$_2$ derived in Appendix
\ref{apph2heating} we obtain

 \begin{eqnarray}\label{h2heat}
 \Gamma_{\mathrm H_2^\star}&=&n_{\mathrm
 H_2}\,\frac{\chi\,P}{1+\left(\frac{A_\mathrm{eff}+
 D_\mathrm{eff}}{\gamma \, n}\right)}\,\Delta E\nonumber \\
 &=&n_{\mathrm
 H_2}\,\frac{9.4\times 10^{22}\,\chi}{1+\left(\frac{1.9\times 10^{-6}+
 4.7\times 10^{-10}\,\chi}{\gamma \, n}\right)}\\
 \label{h2cool}
 \Lambda_{\mathrm H_2}&=&n\,n_{\mathrm H_2}\,\Delta
 E\,\gamma\,\exp(-\Delta E/kT) \nonumber
 \\
 & &\times\frac{A+D}{\gamma\,n+A+D}\nonumber \\
 &=&n\,n_{\mathrm H_2}\,9.1\times 10^{-13}
\,\gamma\,\exp(-6592\,\mathrm{K}/T) \nonumber
 \\
 & &\times\frac{8.6\times 10^{-7}+2.6\times 10^{-11}\,\chi}{\gamma\,n+8.6\times 10^{-7}+2.6\times 10^{-11}\,\chi}
 \end{eqnarray}
with a collisional rate coefficient $\gamma=5.4\times 10^{-13}\,\sqrt{T}$~s$^{-1}$~cm$^{-3}$.
 As the molecular constants do not
depend on the metallicity, only the $Z$-dependence of 
$n\sub{H_2}$ changes the H$_2$ de-excitation heating.

The relative reduction of the H$_2$ heating at high radiation fields
is demonstrated in Figure~\ref{bal_heat}
comparing the PE heating and the H$_2$ de-excitation heating
for the different parameter regimes. We see that at any given density
$\Gamma_{\mathrm PE}^{net}$ exceeds $\Gamma_{\mathrm H_2}^{net}$ beyond a
certain $\chi$ value, but that this limit increases with the gas
density.

Although $\Gamma_\mathrm{PE}$ and $\Gamma_{\mathrm H_2^\star}$ are
the two main heating terms it is necessary to account for a third
process in order to achieve a reasonable approximation of the full
energy balance. For UV fields $\chi< 10^3$, and densities
$n<10^4$ cm$^{-3}$, H$_2$
formation heating may contribute significantly. Assuming that each formation process
releases $1/3$ of its binding energy to heat the gas \citep{SD89}, the corresponding
heating rate is:
\begin{equation}\label{rate_form}
\Gamma_\mathrm{form}=2.4\times10^{-12}\,R\,n\,n_{\mathrm H} \;\;\mathrm{erg\,cm^{-3}\, s^{-1}}
\end{equation}
Summing over all three processes we obtain the total heating rate
\begin{equation}\label{totalheating}
\Gamma_\mathrm{tot}=\Gamma_\mathrm{PE}^{net} + \Gamma_{\mathrm
H_2}^{net} +\Gamma_\mathrm{form}
\end{equation}

\subsubsection{Metallicity dependence}\label{metdep}
\begin{figure}[htpb]
\resizebox{\hsize}{!}{\includegraphics[width=16cm]{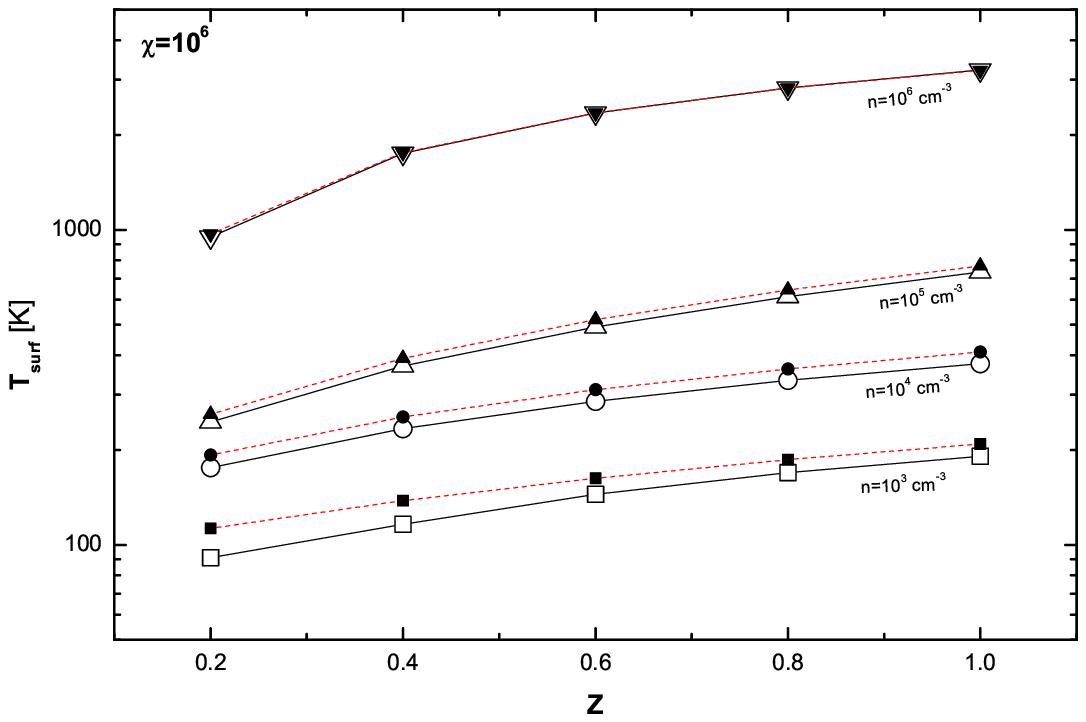}}
\resizebox{\hsize}{!}{\includegraphics[width=16cm]{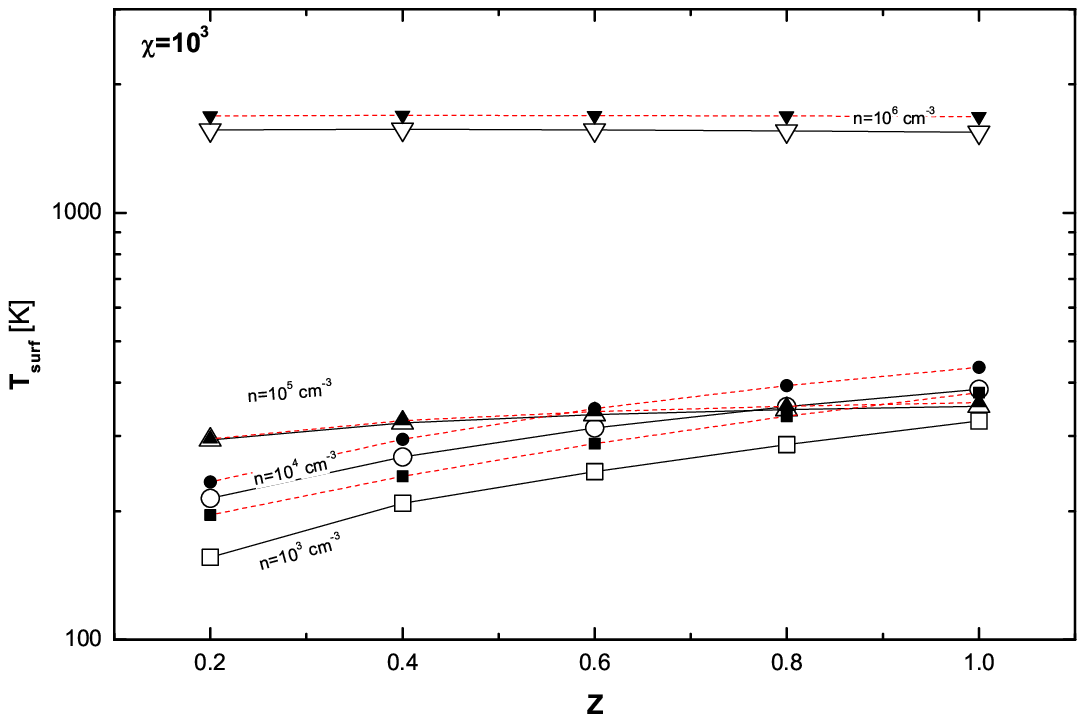}}
\resizebox{\hsize}{!}{\includegraphics[width=16cm]{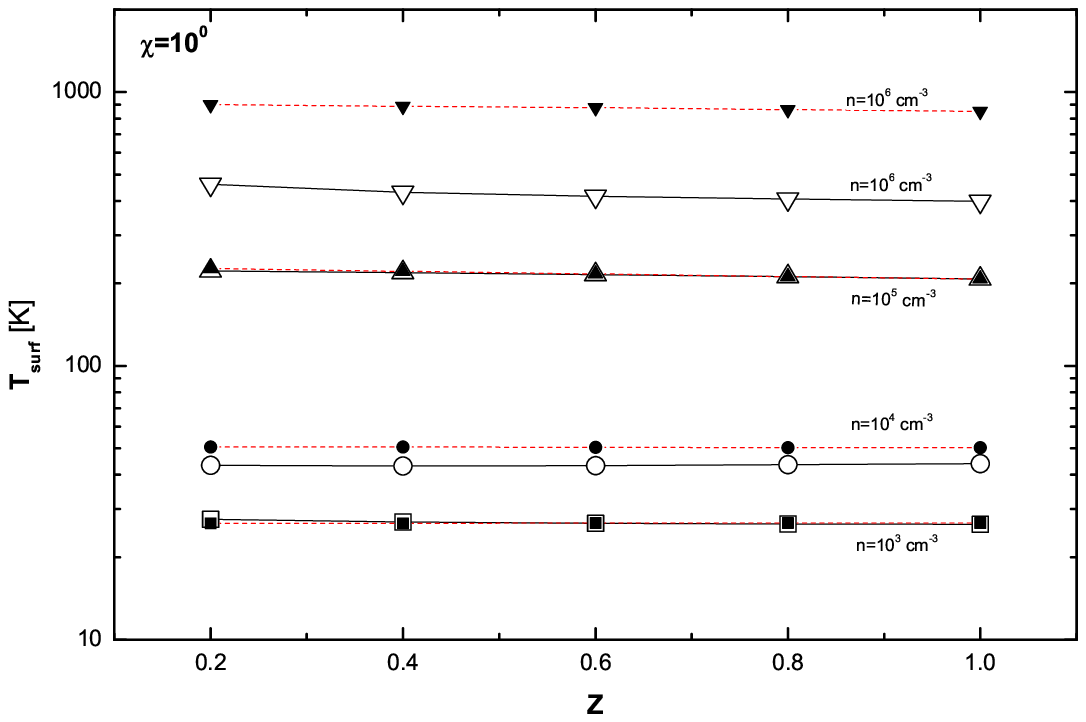}}
\caption{Comparison of the KOSMA-$\tau$ results (open symbols)  and 
the semi-analytic
values (filled symbols) for the surface temperature against the
 metallicity. The top panel a) is for an UV field of $\chi=10^6$, the bottom plot b) is for 
$\chi=10^3$, and the bottom panel c) is for an UV field strength of $\chi=1$. The different symbols indicate different surface densities.
}\label{plot_tsurf1}
\end{figure}

\begin{table}[tbph]
\begin{center}
\begin{tabular}{c|c|c} 
 & low UV field & high UV field\\
 & {\small $\chi < 100$} & {\small $\chi \gg 100 $}\\  \hline & &\\
high &$\Gamma_\mathrm{PE}\sim Z$ &$\Gamma_\mathrm{PE}\sim Z^{\,2}$\\
density &$\Gamma_\mathrm{H_2^*}\sim\frac{Z}{1+2Z}$ &$\Gamma_\mathrm{H_2^*}\sim Z$ \\
{\small$(n\gtrsim 10^6)$} &$\Gamma_\mathrm{H_2-form}\sim{Z}$ &$\Gamma_\mathrm{H_2-form}\sim Z$ \\
& $\Lambda_\mathrm{rec}\sim Z^{1.5}$ & $\Lambda_\mathrm{rec}\sim Z^{1.5}$ \\
& $\Lambda_\mathrm{tot}\sim Z$ & $\Lambda_\mathrm{tot}\sim Z$ \\
& & \\ \hline & &\\
low &$\Gamma_\mathrm{PE}\sim Z$ & $\Gamma_\mathrm{PE}\sim Z...Z^{1.5}$\\
density &$\Gamma_\mathrm{H_2^*}\sim Z$& $\Gamma_\mathrm{H_2^*}\sim Z$\\
{\small$(n\lesssim 10^3)$ }&$\Gamma_\mathrm{H_2-form}\sim{Z}$ &$\Gamma_\mathrm{H_2-form}\sim Z$ \\
&$\Lambda_\mathrm{rec}\sim Z$ &$\Lambda_\mathrm{rec}\sim Z...Z^{1.5}$ \\
& $\Lambda_\mathrm{tot}\sim Z$ & $\Lambda_\mathrm{tot}\sim Z$ \\
&& \\ \hline 
\end{tabular}
\caption{Metallicity dependence of the individual heating processes.}\label{tab1}
\end{center}
\end{table}

An inspection of the heating and cooling functions described above
reveals their metallicity dependence. Table \ref{tab1} summarizes the
scaling relations.
The radiative cooling functions are
linear in $Z$. The photoelectric
heating depends on the metallicity via $\epsilon(Z)\times Z$. The influence of 
$\epsilon(Z)$ can be neglected as long as $n/\chi\gtrsim 100$. Thus, 
for low UV fields $\Gamma_{PE}\sim Z$. For higher values of $\chi$ 
the efficiency accounts for an additional influence due to the electron density
 which is proportional to $Z$ for high densities, and independent of $Z$ 
for very low densities. The metallicity dependence for densities between $10^3...10^5$
cm$^{-3}$ is not trivial.  Eq.(\ref{densityelectron}) shows that the electron 
density is linear in $Z$ for very high $n$. In the intermediate range this dependence
roughly shifts from $Z^0$ to $Z^1$.   
This leads to  $\Gamma_{PE}\sim Z^2$ for high values of $n$ and  $\Gamma_{PE}\sim Z$
for very low densities.

 The recombination cooling depends on the metallicity
through the electron density, resulting in $\Lambda_{rec}\sim Z$ and $ Z^{1.5}$
for low and high densities respectively. 
The $Z$-dependence in the hydrogen heating (Eq. \ref{h2heat} and \ref{h2cool}) 
comes from the hydrogen density
which depends on the metallicity as $Z/(1+2\,Z)$ for high densities and low 
values of $\chi$, and as $Z$ otherwise. 

As a result we show in Fig. \ref{plot_tsurf1} the surface temperature of model clouds 
for a variety of different UV field strengths and densities computed from
the analytic approximation and from the full KOSMA-$\tau$ PDR model. 
We covered a parameter space ranging from $n=10^3\ldots 10^6$ cm$^{-3}$
 and $\chi=10^0\ldots 10^6$. It is obvious that the metallicity dependence varies 
strongly over the parameter space. We obtain a good agreement for low and high UV
fields. Even in the intermediate UV and density range, where the quantitative
accord is weaker, the qualitative dependence of $T_{surf}$ on $Z$ is well
reproduced 
by the semi-analytical model.

The $Z$-dependence of the temperature can be understood by comparing the 
dominant net rates of heating $\Gamma/Z$ and cooling $\Lambda/Z$. At high
UV fields, where the PE heating dominates, the heating is proportional
to $~Z^2$ at high densities. For a density of $n=10^3$ the PE heating is $\sim Z\,\sqrt{Z^2+1}$. 
Due to the high UV field $n/\chi<100$ for all given densities thus the electron density
influences the heating also for small values of $n$. If the density increases the term 
$14.4\,T^{0.75}/n$ vanishes and $\Gamma_{PE}\sim Z^2$. This is reflected
in the slopes of the surface temperature in Fig.~\ref{plot_tsurf1} (top).
For intermediate FUV fields we find a similar behavior with the addition that $n/\chi$ 
is $<100$ for high densities and $\ge 100$ for low densities, thus the metallicity dependence
shifts from $~Z^2$ to $~Z$ with increasing density. This shift can be seen in the middle
plot in Fig.~\ref{plot_tsurf1}.
 
When H$_2$ vibrational de-excitation heating dominates (compare Fig.~\ref{bal_heat}), 
the corresponding rate varies as
$Z/(1+2\, Z)$, hence the surface temperature drops as $1/(1+2\, Z)$. This is shown for $n=10^6$~cm$^{-3}$
in Fig.~\ref{plot_tsurf1} (middle) and for $n\ge 10^4$~cm$^{-3}$
in the bottom panel of Fig.~\ref{plot_tsurf1}  which gives the temperatures for a FUV field strength of $\chi=1$. For
 low FUV fields and low densities the temperature is proportional to $Z$ due to the PE heating as seen 
in Fig.~\ref{plot_tsurf1} (bottom).  

 The offset between the semi-analytical approximation and numerical result 
for $\chi=10^3$ is due to a small contribution of additional cooling processes in that 
parameter range. This increases the total cooling efficiency and hence the temperatures in
the full numerical calculations are smaller. This 
also holds for $n=10^6$~cm$^{-3}$ and $\chi=1$. In that case the cooling is dominated by CO line 
cooling which is stronger than [OI]~63$\mu$m and also by H$_2$O cooling which
is comparable to [OI]~63$\mu$m. Here our initial assumptions are clearly underestimating the 
overall cooling. Even so this does not change the behavior with $Z$ which is well reproduced.

\begin{figure}[t]
\resizebox{\hsize}{!}{\includegraphics[width=17cm]{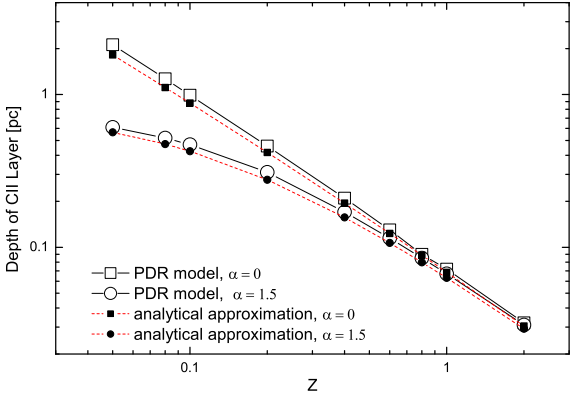}}
\caption{The width of the C$^+$ layer is plotted against the metallicity.
The open circles represent spherical clump of mass $M=10^3$~M$_\odot$, density
$n_0=10^4$~cm$^{-3}$, FUV field $\chi=100$, and the power-law index of the density profile, $\gamma=1.5$,
whereas the open squares represent a spherical clump of mass of $10^{6}$~M$_\odot$
with $\gamma=0$. The filled circles and squares represent our analytical estimates as explained
in Sect.~\ref{ciiw}.
\label{figciiwidth}}
\end{figure}
 As there is some debate on $D/G$ we tested as an extreme example
 $D/G\propto Z^2$ instead of linearity. This mainly changes the 
behavior of the heating rates. The dominant surface cooling processes do not depend on
$D/G$, but only on the elemental abundances, while the heating processes are affected by
an altered $D/G$. This leads to a decreased heating efficiency for $Z<1$, hence the 
surface temperature is significantly lower if we assume $D/G\propto Z^2$. In the 
extreme example of $\chi=10^6$, $n=10^6$~cm$^{-3}$, and $Z=0.2$ we find 
$T_\mathrm{surf}=240$~K, a factor of 4 smaller than for
 $D/G\propto Z$.
The cooling inside the cloud also depends somewhat on $D/G$, since 
the escape probability of cooling lines 
depends on the dust attenuation.

\section{Variation of the C$^+$ layer size with metallicity}
\label{ciiw}

At the surface of the PDR the FUV radiation ionizes almost all of the carbon atoms.
At larger depths the FUV intensity decreases and
carbon recombines and is eventually incorporated into CO molecules. Thus, a PDR clump
can be subdivided into a CO core surrounded by an atomic carbon shell and an outer C$^+$ envelope.
We examine here, the thickness of the C$^+$ envelope as a function of metallicity $Z$.
We define the C$^+$ envelope thickness as the distance
from the cloud surface to the depth where the abundances of C and C$^+$ are equal.

The dominant reaction channels for the formation and destruction of C$^+$ are
\begin{eqnarray*}
\rm C + \nu    &\to&\rm  C^+ + e^-, \\
\rm C^+ + e^-  &\to&\rm  C + \nu \\
&{\rm and}&\\
\rm C^+ + H_2  &\to&\rm  CH_2^+ +\nu.
\end{eqnarray*}
Assuming, that these are the only reactions which influence the C$^+$ abundance,
the balance equation for the abundance of C$^+$ can be written as,
\begin{equation}
\chi\,I\, n\sub{C} = a\sub{C}\, n\sub{C^+}\, n_{e} 
                    + k\sub{C}\, n\sub{C^+}\, n\sub{H_2}
                    \label{eq_c-balance}
\end{equation}
where $I$ is the photoionization rate,
$a\sub{C}$ and $k\sub{C}$ are the recombination and radiative association rate
coefficients. We assume that the photoionization rate is attenuated exponentially as
$\exp(-p\,A_V)$ where the factor $p=3.02$ accounts for the difference in the
opacity between visual and FUV wavelengths. As $A_V$ is determined
by dust extinction it scales linearly with $Z$.
In the KOSMA-$\tau$ model we account for an isotropic FUV field, and
integrate over $4\pi$ ray angles. The ionization rate is then
given by,
\begin{equation}
I=3\times 10^{-10}\,\chi\,\int_1^\infty\frac{\exp(-\,3.02\,A_V\,\mu)}{\mu^2}d\mu
\end{equation}
The integral is the second order exponential integral $E_2(3.02\,A_V)$  where
 $\mu=\cos\Theta$ and $\Theta$ is the angle between the ray and the normal direction.

We have defined the radial point $r\sub{C^+}$ as the location where the abundances
of C$^+$ and C are equal, $n\sub{C^+}=n\sub{C}$. If we neglect the contribution of CO
at $r\sub{C^+}$ then $n\sub{C^+}=X\sub{C}\,n/2$ there. 
Results from the PDR model suggest, that the electron density, 
$n_{e}(Z)\approx 2\, n\sub{C^+}(Z)$, thus $n_{e}(Z) \approx \hat{n}\sub{C}(Z=1)\,Z$.
For the sake of simplicity we chose this expression for $n_e$ rather than the one introduced in 
Eq. \ref{densityelectron} which would introduce an additional temperature dependence.
With ${X}\sub{C}= 1.4\times10^{-4}\,Z$ and molecular hydrogen dominating the gas
density, $n\sub{H_2}=n/2$, we can resolve Eq. \ref{eq_c-balance} for
the density at $r\sub{C^+}$
\begin{equation}\label{ciiwidth}
n(r\sub{C^+})=\frac{3\times 10^{-10}\,\chi\, E_2\left(3.02 \,A_V(r\sub{C^+})\right)}
{a\sub{C}\,Z\,1.4\times10^{-4} + 0.5\,k\sub{C}}
\end{equation}
\begin{figure}[htb]
\resizebox{\hsize}{!}{\includegraphics[width=17cm]{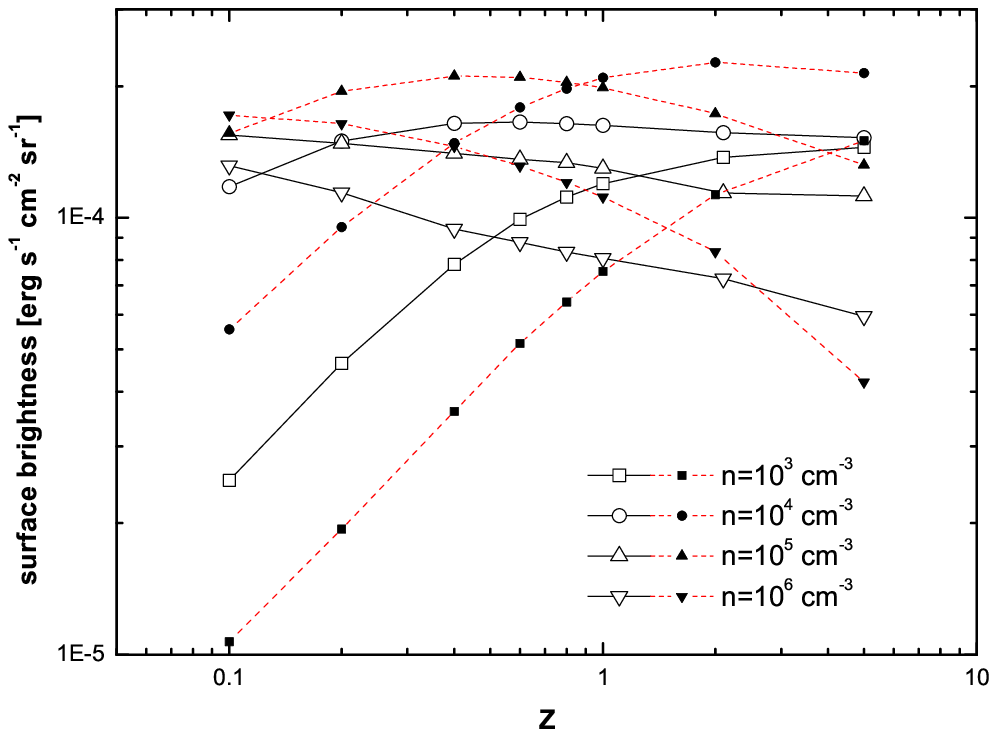}}
\caption{
The surface brightnesses of the [CII]~158$\mu$m (dashed-dotted)
emission lines from the KOSMA-$\tau$ results (open symbols) and
the approximation from Eq.(\ref{surfbright}) (filled symbols)
plotted against the metallicity for  different densities. 
The cloud mass is  §M=10§~M$_\odot$, the UV field strength is  $\chi=100$, the
assumed central temperature is $T_c=35$~K, and $\lambda=4$.
\label{icii}
}
\end{figure}
\begin{figure}[htb]
\resizebox{\hsize}{!}{\includegraphics[width=17cm]{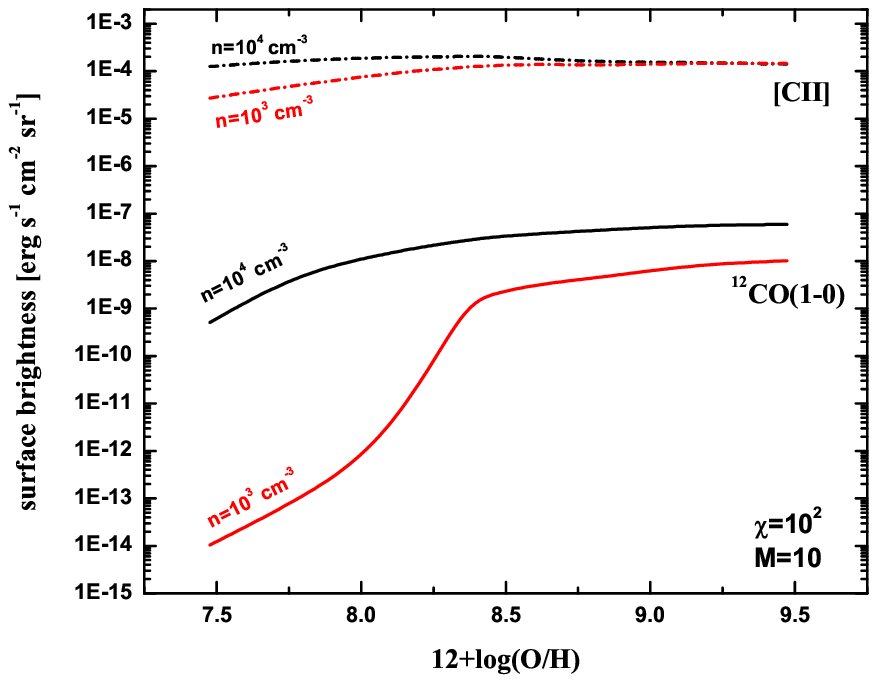}}
\caption{
The surface brightnesses of the [CII]~158$\mu$m (dashed-dotted)
and $^{12}$CO(1-0) 2.6~mm (solid)
emission lines 
plotted against the metallicity for two different densities. 
The cloud mass is  $M=10$~M$_\odot$ and the UV field strength is  $\chi=100$.
\label{ciico}
}
\end{figure}

For a given radial density distribution $n(r)$ and FUV field $\chi$,
Eq.~\ref{ciiwidth} can be numerically solved to obtain $r\sub{C^+}$, 
 or correspondingly the width of the $\mathrm C^+$ layer $D\sub{C^+}=R-r\sub{C^+}$.
In Fig. \ref{figciiwidth} we compare the results from this equation with
detailed PDR model calculations using the KOSMA-$\tau$ model
for $n_0=10^4~cm^{-3}$ and $\chi=100$. Here two types of 
density structures are used:
(a) $n(r)=n_0\,(r\,/R)^{-\gamma}$ for $0.2\,R\,\le\, r\, < R$, $n(r)=0$ for $r>R$, 
and $n=n_0\,0.2^{-\gamma}$ for $r < 0.2R$, with the total cloud radius $R$;
(b) a constant density,
$n(r)=n_0$, or equivalently $\gamma=0$. 

\begin{figure*}[bt]
\includegraphics[width=17cm]{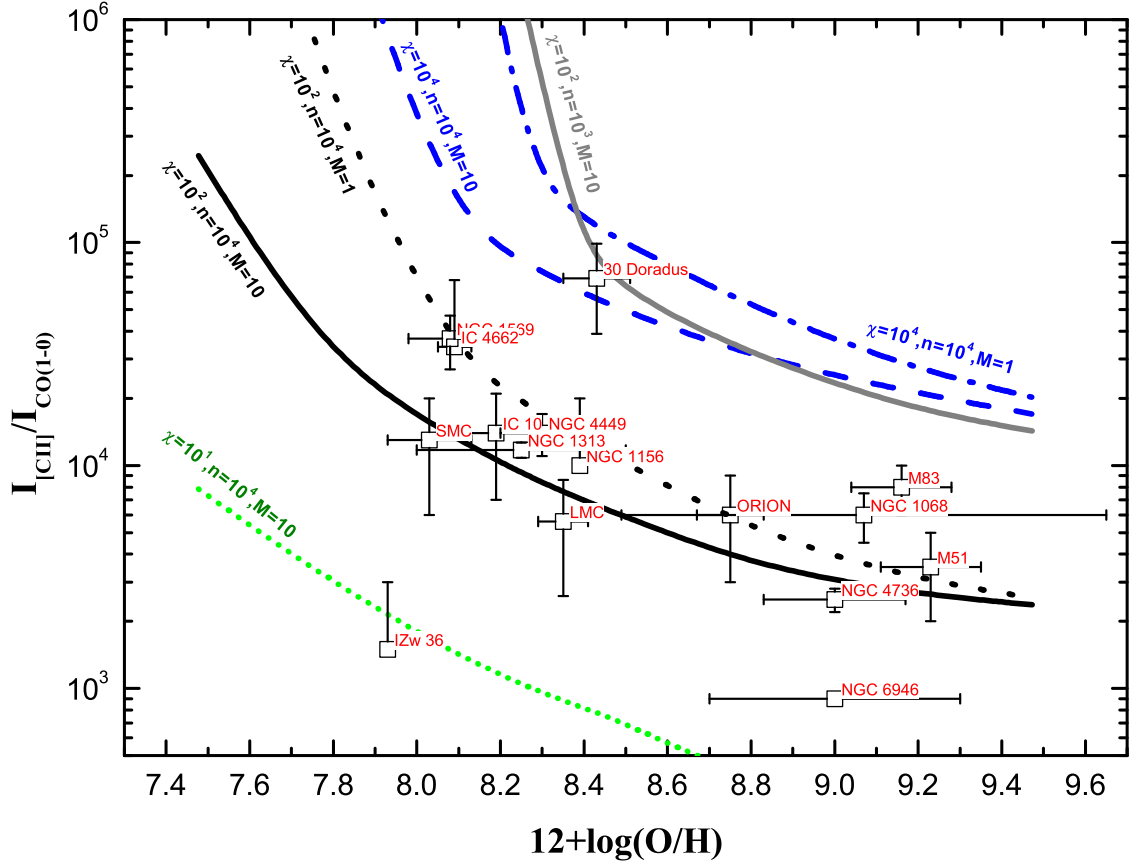}
\caption{
The intensity ratio $I_{\mathrm{[CII]}158\mu m}$/$I_{\mathrm{CO(1-0)}}$
is plotted against the metallicity ($Z=1$ is equivalent to 12+log(O/H)=8.48). The lines denote the KOSMA-$\tau$ results for
two different clump masses $M=1$ and $M=10$~M$_\odot$.  The density at the surface of the
clumps and the UV field strength in units of the standard Draine field are given in the plot.
 The observed ratios of nearby galaxies are plotted
as squares. 
\label{clumpyplot}
}

\end{figure*}
The squares representing the constant density model show that the C$^+$
layer width depends approximately as $Z^{-1.1}$ on the metallicity which
closely matches with the assumption of an inverse proportionality by \citet{bolatto99}.
The circles in Fig.~\ref{figciiwidth} represent the model by the 
 $\gamma=1.5$. They can be numerically fitted with
function, $D\sub{C^+}=(1.195+5.758 Z)^{-1.47}$~pc. Fig.~\ref{figciiwidth} shows that our results using
Eq.~\ref{ciiwidth} agrees well with the model calculations. However the estimated
widths are slightly higher than the model calculations, which reflects the fact that
there are more reactions which quantitatively influence the chemistry of C$^+$.
\begin{table*}
\caption{Metallicities and observed [CII]/CO(1-0) line ratios of nearby galaxies and Galactic star
forming regions. The derived values for gas density and FUV strength are also given 
if available.\label{tab_ratio}}
\centering
\begin{tabular}{c c c c}
\hline \hline
Object & 12+log(O/H) &[CII]/CO(1-0)  & References\\
\hline
NGC 1068 & 9.07&6000  &14,2,12\\
NGC 1156 & 8.39&10000 &4\\
NGC 1313 & 8.25&11740 &14,7\\
NGC 1569 & 8.08&37000 &13,4,6\\
NGC 4449 & 8.3 &14000 &8,6\\
NGC 4736 & 9   & 2500 &14,2,12,10\\
NGC 6946 & 9   &  900 &14,12\\
   IC 10 & 8.19&14000 &14,8,6\\
  IZw 36 & 7.93&$<3000$ &8\\
     M83 & 9.16& 8000 &14,2,5\\
     M51 & 9.23& 3500 &14,9,5\\
 IC 4622 & 8.09&34000 &3,4\\
   Orion & 8.75& 6000 &2,11\\
     LMC & 8.35& 5600 &1\\
30 Doradus &8.43&69000&1\\
     SMC & 8.03&13000 &1\\
\hline
\end{tabular}
\begin{list}{}{}
\item[References.---](1) Bolatto et al. 1999; (2) Crawford et al. 1985; (3) Heydari-Malayeri et al. 1990;
(4) Hunter et al. 2001; (5) Kramer et al. 2005 (6) Lord et al. 1995; (7) Luhmann et al. 2003; (8) Mochizuki et al. 1998;
(9) Nikola et al. 2001; (10) Petitpas\&Wilson 2003; (11) Simon et al. 1997 (12) Stacey et al. 1991; (13) Talent 1980;
(14) Zaritsky et al. 1994.
\end{list}
\end{table*}

\section{[CII] emission as a function of $Z$}
\label{clumpysection}

The [CII] emission as well as the [CII]/CO(1-0) line ratio is typically considered to
be a good tracer of star formation \citep{stacey91}. 
The intensity ratio $I_{\mathrm{[CII]}158\mu m}$/$I_{\mathrm{CO(1-0)}}$, observed in many nearby
low metal galaxies, is higher than for sources with solar and super-solar metallicities 
\citep{madden97, mochi98, bolatto99, madden00, hunter01}. Table~\ref{tab_ratio} summarizes
available line ratios and metallicities of nearby galaxies. The corresponding numbers
for Orion are also given as a Galactic reference.

This dependence has been modelled by
\citet{bolatto99} assuming that the size of the C$^+$ region scales inversely with
metallicity, and assuming a constant temperature for the gas.

For our spherical PDR model we compute the surface brightness as function of 
clump mass/radius and metallicity. The surface brightness is the projected average intensity
\begin{equation}
\bar{I}=\frac{2\,\pi\,\int_{\,0}^{\,R}I(p)p\,dp}{\pi\,R^2}
\end{equation}
of the spherical clump, where $I(p)$ is the specific intensity along a ray with impact 
parameter $p$ \citep{stoerzer96}. The line intensities depend on the thermal and chemical structure of the cloud. 
Additionally, for spherical clouds an effective area-filling factor of the
emissive region has to be considered. Particularly lines, which are formed in central regions of the cloud are 
influenced by this area-filling factor. A good example is the surface brightness of
$^{12}$CO(1-0). For a density of n=$10^3$~cm$^{-3}$ and low metallicities almost the whole
cloud is devoid of CO due to photodissociation. A higher density or
metallicity results in a  
larger
CO core and a higher surface brightness. For species which are mainly emitting at the surface of the cloud
this filling effect is negligible. 

In the prior sections we derived approximate expressions for the surface temperature of a PDR as well as 
for the expected depth of the C$^+$ envelope. We can use these approximations to estimate the total [CII] surface
 brightness of the PDR. The local emissivity $\Lambda_\mathrm{[CII]}(r)$ from Eq.(\ref{cii_cooling}) 
can be used to calculate the line integrated 
intensity of the PDR in the optically thin case
\begin{equation}\label{intint}
I_{int}=\int_{r_{\mathrm C^+}}^R 4\,\pi\,r^2\,\frac{\Lambda_\mathrm{[CII]}(r)}{4\,\pi} dr\;\; \mathrm{erg\, s^{-1}\, sr^{-1}}
\end{equation}
with the total radius R. From Eq.~(\ref{intint}) follows the mean surface brightness
of the cloud
\begin{equation}\label{surfbright}
\bar{I}_{\mathrm{[CII]}}=\frac{I_{int}}{\pi\,R^2}\;\;\; \mathrm{erg\, s^{-1}\, cm^{-2}\,sr^{-1}}
\end{equation}
To calculate $I_{\mathrm{[CII]}}$we assume an exponential temperature profile $T(r)=T_c+T_\mathrm{surf} E_2(\lambda \,A_V)$ with  
the the second order exponential integral $E_2$, and an arbitrary fitting parameter $\lambda$. 
The central temperature $T_c$ and the parameter $\lambda$ are different for each
set of PDR parameters. For demonstration purposes we chose $T_c=35$~K and $\lambda=4$ to estimate the surface brightnesses for multiple PDRs simultaneously. In Fig.~\ref{icii} we compare these approximations with the detailed KOSMA-$\tau$ results for I$_\mathrm{[CII]}$. The metallicity dependent behavior is reproduced very well and the quantitative agreement is within a factor of 2 assuming the same temperature profile for all 4 models! If we drop this assumption and use individual temperature profiles for each model the agreement is 10-30\%.

Importantly, the total surface brightness does not scale linearly with the surface density of the clouds. Rather it peaks for intermediate values of $n$, depending on $Z$. 
This is mainly a geometrical effect, which can be understood by some qualitative arguments.  If we assume that [CII] is optically thin, we see all $\mathrm C^+$ atoms. In the low density case, where ionized carbon fills the whole cloud, the surface brightness is then proportional to $n\,(V/A)$, with the volume of the cloud $V$ and the projected area $A$, hence $\bar I_\mathrm{[CII]}\propto n\,R$. But $R\propto n^{-1/3}$, since we kept the cloud mass constant, and thus we find  $\bar I_\mathrm{[CII]}\propto n^{2/3}$. For higher densities the width of the  $\mathrm C^+$ layer decreases faster than it is compensated by the growing $n$. The relative thickness of the $\mathrm C^+$ layer becomes very small for higher densities (i.e. if $D_\mathrm{C^+} << R$). The surface brightness then is proportional to $n\,4\,\pi\,R^2\,D_\mathrm{C^+}/(\pi\,R^2)$, hence  $\bar I_\mathrm{[CII]}\propto n\,D_\mathrm{C^+}$. We observe a reduced surface brightness caused by the geometry of the cloud. This is inverse to the common area filling effect for optically thick lines, like e.g. CO (1-0), where the projected area of the CO core decreases with decreasing density as demonstrated in Fig.~\ref{icii}. This was also mentioned by \cite{stoerzer96}. This means, that even though the local emissivity $\Lambda_\mathrm{[CII]}$ scales linearly with $n$ (see Eq.~\ref{cii_cooling}), this is not true for the total surface brightness. As a second order effect we also notice a temperature dependence of the local emissivity. The differences between our analytical model and the detailed temperature structures from the PDR calculations are responsible for most of the deviations shown in Fig.~\ref{icii}.

We use our PDR model calculations to study the
metallicity dependence of the [CII]/CO(1-0) line ratio. We adopt a density of $10^4$~cm$^{-3}$, a $\gamma=1.5$, and a UV field,
$\chi=10^2$ similar to the values assumed by \citet{bolatto99}. 
Their prediction is an average over a clump
ensemble as a model for the large scale emission from the ISM. In
contrast, we start here
by investigating the metallicity dependence for a single, typical
clump. The discussion below shows that this is already sufficient to reproduce
the observed trends versus metallicity. The detailed investigation of the
effects of averaging over a clump ensemble are left to a subsequent paper.
Fig.~\ref{ciico} summarizes the results for this typical clump. 
The dominance of the geometrical effect
is reflected in the almost constant [CII] surface brightnesses in Fig.~\ref{ciico}.
Hence the line ratio [CII]/CO decreases for increasing metallicities and tends to be constant for very
high values of $Z$.     
   
Figure~\ref{clumpyplot} shows that
the trend in the observed ratios in normal galaxies can be represented by a single-clump model with $M = 1 \ldots 10^1$~M$_\odot$, 
 shown as solid and dotted lines in Fig.~\ref{clumpyplot}. We also plotted model results 
for different cloud parameters to demonstrate how different observations may
be explained by different local physical conditions, e.g. the higher observed 
ratio for the 30 Doradus region can be explained by a similar model,
but exposed to an UV field of  $\chi=10^4$ (or alternatively by a clump of less mass). This is consistent
with derived FUV strengths for 30 Doradus \citep{kaufman99}.
The peculiar source IZw 36, with a very low [CII]/CO ratio at extremely low
metallicity can be approximated by a model with a lower FUV field of $\chi\approx10$, consistent 
with estimations by \citet{mochi98}.

Our model results for single
clumps reproduce qualitatively the results shown by the semi-analytical clumpy model
of \citep{bolatto99}
for a clump ensemble in reproducing the trends versus metallicity. Thus
we can confirm Bolatto's findings whan taking the detailed physical and
chemical structure of the clumps into account. 

From a practical point of view it is obvious that a clumpy ensemble of
different clouds should be closer to the true local conditions than a single
spherical clump, but we find that a clumpy approach is not necessary to 
explain the observerved trend with $Z$. To
model the [CII]/CO line ratio of a particular source in detail it may of
course be necessary to apply a clumpy approach. But to understand the general
behavior for different metallicities it is sufficient to 
consider a single, typical clump.
\section{Summary}

We study the effects of  metallicity variations on the gas temperature
and [CII] emission line properties of spherical PDRs. We find that
the surface temperature of PDRs at high UV fields varies linearly with metallicity.
For low UV fields and high densities this metallicity behavior of the surface temperature is converse,
 showing an inverse dependence with metallicity due to the dominant H$_2$ heating. We
introduce a new two level FUV H$_2$ heating and cooling function that properly accounts 
for energy losses via vibrational collisional excitations. 

We examine the dependence of the C$^+$ envelope on metallicity and find that
its geometrical depth scales 
inversely with $Z$. 
This produces 
a higher [CII]/CO(J=1-0)
line ratio at lower metallicities.  We used the numerical results from the
KOSMA-$\tau$ model to study the dependence of PDR emission lines with
metallicity.  The observed variation of [CII´]/CO(J=1-0) with metallicity can
be explained 
well by a single-clump model and
it is not necessary to refere to an average over a clump ensemble. 
We conclude that the [CII]/CO(J=1-0) line ratios for sources with
differing metallicities do not provide a strong constraint on the clumpy
morphology of a molecular clouds.

\begin{acknowledgements}
This work is supported by the Deutsche Forschungs Gemeinschaft
(DFG) via Grant SFB\,494. AS thanks the Israel Science 
Foundation for support. We thank the anonymous referee for her/his
helpful comments.
\end{acknowledgements}

\clearpage

\appendix
\onecolumn

\section{Cooling functions}\label{appcooling}

The cooling of the gas is dominated by fine structure line
emission of [CII] and [OI]. The line cooling rate can always be written as
$\Lambda_{ul}=n_u\, A_{ul}\, E_{ul}\, \beta(\tau_{ul})$ erg~s$^{-1}$
cm$^{-3}$ where $\beta$ is the escape probability, $A_{ul}$ is the
transition probability and $n_u$ the number of atoms in the upper
state $u$ and $E_{ul}$ is the corresponding transition energy
\citep{HM79}. Below the critical density for the [OI] emission
$n_\mathrm{cr}=8.5\times 10^{-5}\,(100K/T)^{0.69}$ cm$^{-3}$, the
main cooling is provided by the 158 $\mu$m [CII] line. The general
cooling rate of a two-level system S can be expressed as:
\begin{equation}
\Lambda_{ul}=\frac{n_{\mathrm S}\, A_{ul}\, E_{ul}\, \beta}{1+
\frac{g_l}{g_u}
\exp(E_{ul}/kT)(1+\frac{n_\mathrm{cr}\,\beta}{n})}\;\mathrm{erg\,cm^{-3}\,
s^{-1}}
\end{equation}

The critical density for collisional de-excitation
$n_\mathrm{cr}=A_{ul}/\gamma_{ul}$ for the [CII] 158 $\mu$m
transition is $2.6\times10^3$ cm$^{-3}$. The densities we are
interested in are in the range of $10^3...10^6$ cm$^{-3}$. 
Inserting the numerical values for [CII] and
assuming a relative carbon abundance of $1.4\times 10^{-4}\times
Z$ we obtain the cooling rate for the [CII] 158 $\mu$m
transition:
\begin{equation} \label{cii_cooling}
\Lambda_\mathrm{CII}=\frac{2.02\times 10^{-24}\, n\, Z}{1+
\frac{1}{2}
\exp(92/T)(1+\frac{1300}{n})}\;\;\mathrm{erg\,cm^{-3}\, s^{-1}}
\end{equation}

The general cooling rate of the [OI]$63\mu$m ($^3P_1\to ^3P_2$) and [OI]$146\mu$m ($^3P_0\to ^3P_1$) transitions, only
accounting for transitions between neighboring levels, are:

 \begin{eqnarray}
\Lambda_{12}&=&A_{12}\, E_{12}\,\beta\, Z \,\left(\frac{n_\mathrm{OI}\, \exp(E_{01}/T)\,g_1\,n\,(n+\beta\,n_{cr,01})}{g_0\,n^2\,´\,\exp(E_{01}/T)\,(n+\beta\,n_{cr,01})\,(g_1\,n\,+\exp(E_{12}/T)\,g_2\,(n+\beta\,n_{cr,12}))}\right)\;\;\;\mathrm{erg\,cm^{-3}\, s^{-1}}\\
\Lambda_{01}&=&A_{01}\,E_{01}\,\beta\, Z\, \left(\frac{n_\mathrm{OI}\,g_0\,n^2}{g_0\,n^2\,´\,\exp(E_{01}/T)\,(n+\beta\,n_{cr,01})\,(g_1\,n\,+\exp(E_{12}/T)\,g_2\,(n+\beta\,n_{cr,12}))}\right)\;\;\;\mathrm{erg\,cm^{-3}\, s^{-1}}
 \end{eqnarray}

Inserting the numerical values for [OI] and a relative oxygen
abundance of $3\times 10^{-4}\times Z$ in the above equation leads
to:
\begin{eqnarray}\textstyle
\Lambda_{63\mu m}&=&3.15\times 10^{-14}\,8.46\times 10^{-5}\,\frac{1}{2}\,Z\times\nonumber \\
&&\frac{3\times10^{-4}\,n\,\exp(98\,\mathrm{K}/T)\,3\,n\,\left(n+\frac{1}{2}\frac{1.66\times 10^{-5}}{1.35\times10^{-11}\,T^{0.45}}\right)}{n^2\,+\,\exp(98\,\mathrm{K}/T)\left(n\,+\,\frac{1}{2}\frac{1.66\times 10^{-5}}{1.35\times10^{-11}\,T^{0.45}}\right)\left(3\,n\,+\,\exp(228\,\mathrm{K}/T)\,5\,\left(n\,+\,\frac{1}{2}\frac{8.46\times 10^{-5}}{4.37\times10^{-12}\,T^{0.66}}\right)\right)}\;\;\;\mathrm{erg\,cm^{-3}\, s^{-1}}\\
\Lambda_{146\mu m}&=&1.35\times 10^{-14}\,1.66\times 10^{-5}\,\frac{1}{2}\,Z\times\nonumber\\
&&\frac{3\times10^{-4}\,n\,n^2}{n^2\,+\,\exp(98\,\mathrm{K}/T)\left(n\,+\,\frac{1}{2}\frac{1.66\times 10^{-5}}{1.35\times10^{-11}\,T^{0.45}}\right)\left(3\,n\,+\,\exp(228\,\mathrm{K}/T)\,5\,\left(n\,+\,\frac{1}{2}\frac{8.46\times 10^{-5}}{4.37\times10^{-12}\,T^{0.66}}\right)\right)}\;\;\;\mathrm{erg\,cm^{-3}\, s^{-1}}
\end{eqnarray}
This leads to the total [OI] cooling rate $\Lambda_\mathrm{OI}=\Lambda_{63\mu m}+\Lambda_{146\mu m}$. In the high density case gas-grain collisional cooling is
also contributing to the total cooling of the cloud:
\begin{equation}
\Lambda_\mathrm{g-g}=3.5\times 10^{-34}
\sqrt{T}\left(T-T_\mathrm{grain}\right)\, n^2\, Z \;\; .
\end{equation}
$T_\mathrm{grain}$ is the dust temperature as given by
\citet{HTT91}. The total cooling rate is the sum of the above
individual rates
\begin{equation}
\Lambda_\mathrm{tot}=\Lambda_\mathrm{CII}+\Lambda_\mathrm{
OI}+\Lambda_\mathrm{g-g}
\end{equation}

\section{The electron density at PDR surfaces}\label{appelectron}
\begin{figure*}[h!]
\begin{minipage}[t]{8.5cm}
\includegraphics[width=8.5cm]{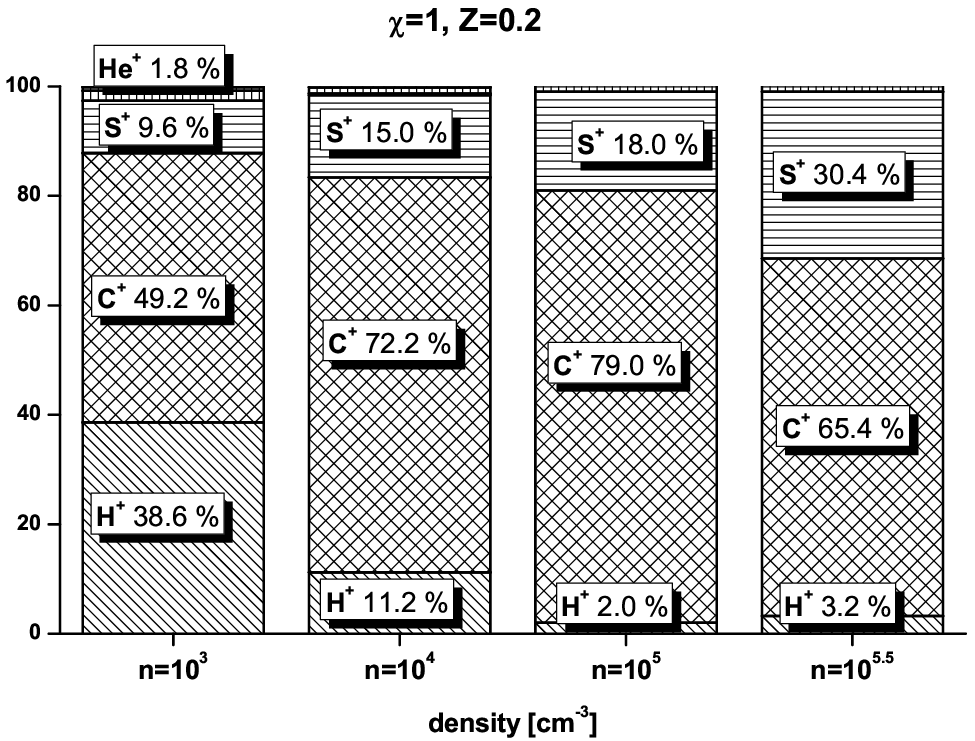}
\end{minipage}
\hfill
\begin{minipage}[t]{8.5cm}
\includegraphics[width=8.5cm]{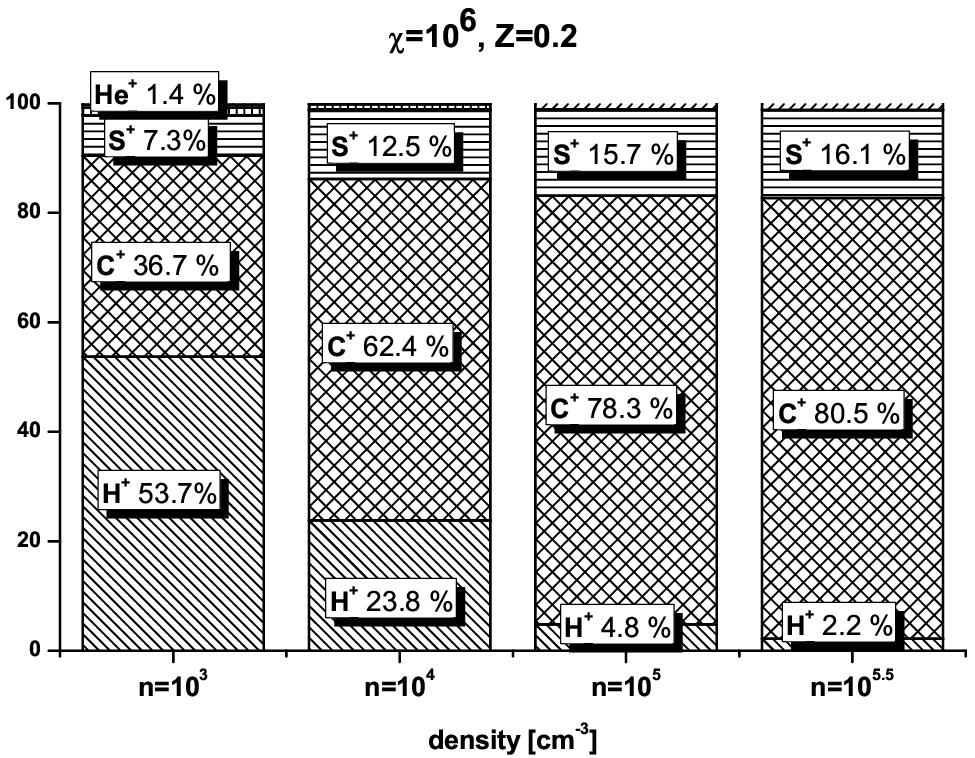}
\end{minipage}
\caption{\small The relative contribution of the various electron donors to the total
electron density $n_e$ at the cloud surface as computed
in the KOSMA-$\tau$ model. The percentages are 
given for different values of surface density $n$ (columns) at FUV fields $\chi=1$ (left) and  $\chi=10^6$ (right)
and a fixed metallicity of $Z=0.2$.
\label{histoplot}
}
\end{figure*}

At the surface of a PDR atomic carbon and sulfur are ionized
by the impinging, unshielded FUV radiation and  atomic hydrogen is ionized 
by cosmic rays (FUV ionization of H is prevented by the Lyman limit). Electrons 
from dust are negligible due to the small number density of dust grains. 
The relative contribution from these electron donors to the
total electron density $n_e$ is shown in Fig.~\ref{histoplot} for the low (left) and
high FUV (right) case. The values are computed with the KOSMA-$\tau$ PDR model. The
histogram shows the increasing importance of atomic carbon as main electron source with
increasing density. Nevertheless the additional contributions of atomic hydrogen and
sulfur are not negligible. Even at very high UV fields, $\chi \approx 10^6$, a
relevant fraction of the electrons is generated in the additional
ionization processes. At gas densities $\sim 10^{5.5}$~cm$^{-3}$ still
16-20\% of the electrons stem from the ionization of atomic
hydrogen and sulfur. This fraction increases rapidly with
decreasing density, with the result that at a number density of
$\sim 1000$~cm$^{-3}$ only $< 40\%$ of the electrons are due to
ionization of atomic carbon. 
\begin{figure*}[h]

\includegraphics[width=17cm]{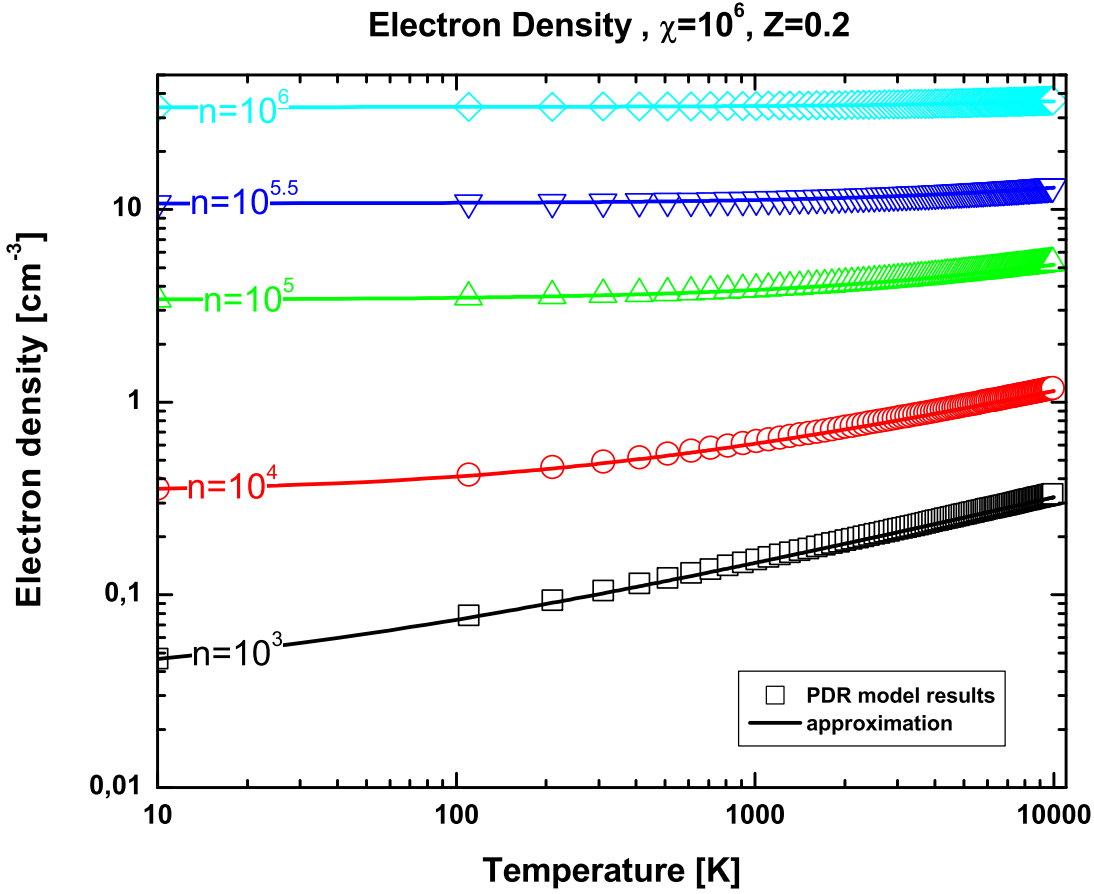}
\caption{\small
A comparison of numerically obtained electron densities (symbols) with 
the analytical expression (lines) in Eq. \ref{densityelectronA},
 at A$_V$=0, $\chi=10^6$, $Z=1$ for
different surface densities.
\label{edenplot}
}
\end{figure*}
We present an analytic approximation to the electron density at
PDR surfaces
which fits the actual behavior quite accurately and which allows
an easy interpretation of the metallicity dependence. We assume  
\begin{equation}\label{balanceE}
    n_{e}=(X_\mathrm C\,+\,X_\mathrm S)\,n\,Z+n_\mathrm{H^+}
\end{equation}
$X_\mathrm C$ and $X_\mathrm S$ are the relative elemental abundances
of carbon and sulfur ( $X_\mathrm C=1.4\times 10^{-4}$, 
$X_\mathrm S=2.8\times 10^{-5}$, \cite{HT99,federman93}).
Together with the balance equation for hydrogen:
\begin{equation} \label{balanceH}
n_{\mathrm H}\,\zeta=a_{\mathrm H} \,n_\mathrm{H^+}\, n_e\,\;\;,
\end{equation}
where $\zeta=2.3\times10^{-17}$~s$^{-1}$ is the ionization
rate due to cosmic rays \citep{SD95}, and
$a_{\mathrm H}=3.5\times10^{-12}(T/300 K)^{-0.75}$~s$^{-1}$ is
the recombination rate, we get the following expression for the
electron density:
\begin{equation}\label{density1}
 n_e=\frac{X}{2}\,n\,Z\,\left[1+\sqrt{1+\frac{T^{0.75}}{n\,Z^2}
\left(\frac{6.05\times10^{-4}}{X}\right)}\,\right]\;\;\;\;{\rm cm^{-3}}
\end{equation}
{\bf with $X=X_\mathrm C\,+\,X_\mathrm S$.}
This is in reasonable agreement with the numerical results
from the detailed PDR calculations but deviates by $5-10\%$. The deviations are due to the different net recombination
rates when considering all ionized species. To account for this changed
rate we have fitted the parameter $a$ in Eq. (\ref{balanceH}) to match
the electron density from the full PDR model.
Assuming $n_\mathrm{H^+}\approx C\,n_\mathrm H\,
\zeta/n_e\,a_{\mathrm H}$ and fitting the constant
$C$ to the numerical values we obtain
\begin{equation}\label{densityelectronA}
 n_e=\frac{X}{2}\,n\,Z\,\left[1+\sqrt{1+\frac{T^{0.75}}{n\,Z^2}
\left(\frac{7.52\times10^{-4}}{X}\right)}\,\right]\;\;\;\;{\rm cm^{-3}}
\end{equation}
The agreement with the KOSMA-$\tau$ results is shown
in Fig.\ref{edenplot}. Please note that in the high radiation case
all C and S at the surface are ionized so that the ratio of electrons 
contributed by them directly reflects their abundance ratio.

\section{H$_2$ vibrational heating}\label{apph2heating}
\begin{figure*}[th!]

\includegraphics[width=16cm]{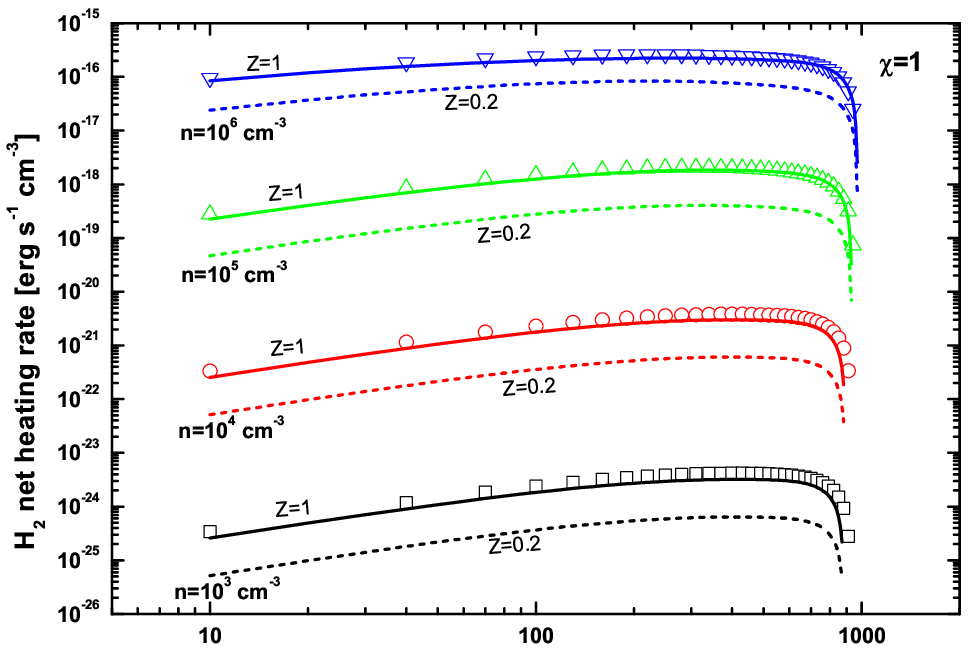}
\vspace{\fill}
\includegraphics[width=16cm]{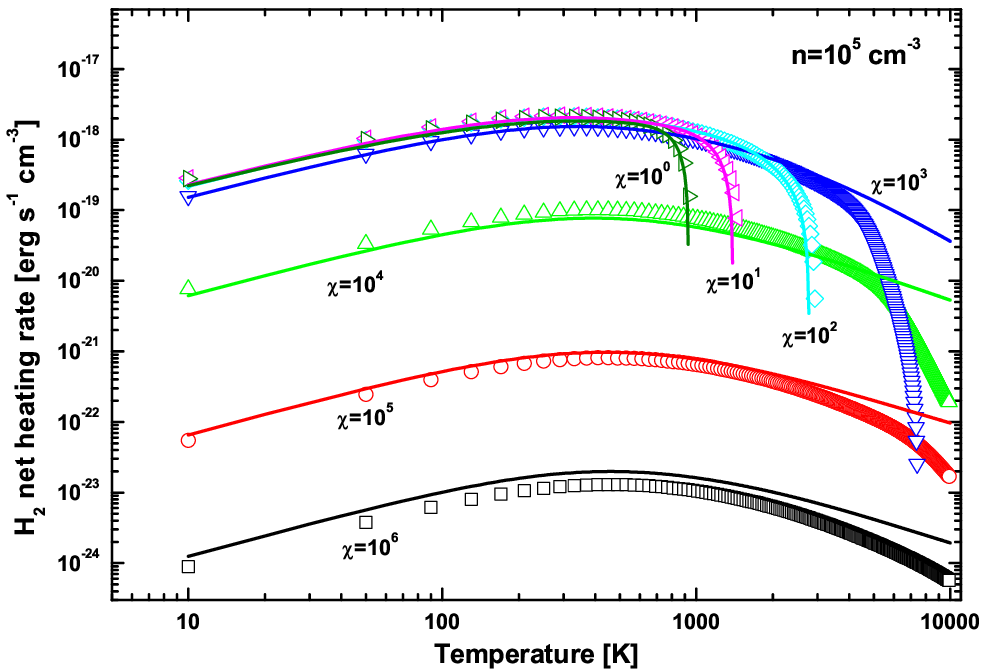}
\caption{\small
The net H$_2$ heating rate is plotted for different values 
of surface density $n$ at a $\chi=1$ FUV field (top) and for different UV field strengths
for a fixed density of $n=10^3$~cm$^{-3}$ (bottom) over the temperature. The symbols
 are the numerical results from
the 15-level system in the KOSMA-$\tau$ model,
the curves show the results using our effective two-level approximation. For 
 $\chi=1$ we  also plot a low $Z$ case (solid, $Z=1$, and dashed, $Z=0.2$). 
The deviations at the highest temperatures and radiation fields are due to
cooling transitions from higher vibrational levels in this regime which
are ignored in our approximation.
\label{h2hn30}
}
\end{figure*}

An important heating processes in dense PDRs is collisional
deexcitation of FUV-pumped H$_2$ molecules \citep{SD95}. 
Here we present a two level approximation for the H$_2$ vibrational
heating and cooling valid in the parameter range where the process
plays a major role (see Sect. \ref{sect_heating}). The approximation
reproduces the net heating rate computed by SD95 assuming transitions
among all 15 vibrational levels in the ground electronic state, but neglecting the
rotational structure.

Vibrational cooling reduces the net heating
at large gas temperatures (see Fig.~\ref{h2hn30}, bottom).
The vibrational cooling is most effective at low $\chi$
for which a large H$_2$ density is maintained.  With PDR temperatures
of typically less than 2000~K (see Sect. \ref{metalsection})
and the energy gap between the two lowest vibrational
levels $\Delta E_{0,1}=5988$~K, we can assume that most of the H$_2$
is always in the ground ($v=0$) level in this regime. Vibrational cooling
is thus basically given by collisional excitation to $v=1$ followed
by either radiative decay or photodissociation.
Using the molecular constants for the lowest vibrational transition
we obtain the collisional cooling rate
\begin{equation}\label{b1}
\Lambda_\mathrm{H_2}=-\Delta E_{1,0}\,\gamma_{1,0}
\exp\left(\frac{-\Delta
E_{1,0}}{k\,T}\right)\,n\,n_\mathrm{H_2}\,
\frac{A_{1,0}+\chi\,D_1}{\gamma_{1,0}\,n+A_{1,0}+\chi\,D_1}
\end{equation}
with the spontaneous emission rate coefficient $A_{1,0}=8.6\times 10^{-7}$~s$^{-1}$,
the collisional rate coefficient
$\gamma_{1,0}=5.4\times 10^{-13}\sqrt{T}$~s$^{-1}$~cm$^{-3}$,
and the standard photodissociation rate for the $v=1$ level of
$D_{1}=2.6\times 10^{-11}$~s$^{-1}$ \citep{SD95}. We found that the fit can be improved
if  $\Delta E_{0,1}$ is increased by 10\%.

In contrast vibrational heating is important when the FUV radiation
field provides a significant pumping  to higher vibrational states.
Thus we define a separate equivalent two-level system for the
heating. It is characterized by the effective coefficients
$\Delta E_\mathrm{eff}$, $A\sub{eff}$, $\gamma\sub{eff}$, and $D\sub{eff}$
providing the same heating rate as the full 15 level system
\begin{eqnarray}\label{b2}
\Gamma_\mathrm{H_2^*}&=&n_\mathrm{H_2}\sum_j\sum_{i\ge
j}\frac{\chi\,P_i\,\Delta E_j}{1+[A_j+\chi\,D_j]/[\gamma_j\, n]}\\
\label{h2eff}
&=&n_\mathrm{H_2}\frac{\chi\,P_\mathrm{tot}\,\Delta
E_\mathrm{eff}}{1+[A_\mathrm{eff}+\chi D_\mathrm
{eff}]/[\gamma_\mathrm{eff}\, n]}
\end{eqnarray}
The quantity $P_i$ denotes the formation rate of vibrationally
excited H$_2$ for the different levels, $P\sub{tot}$ represents
the sum rate over all levels. 
The effective coefficients can be easily obtained by considering
different asymptotic values of the density $n$ and the radiation
field $\chi$. This yields
$P_\mathrm{tot}\cdot\Delta E_\mathrm{eff}=9.4\times 10^{-22}$~erg~s$^{-1}$,
$\gamma_\mathrm{eff}=\gamma_{1,0}, D_\mathrm{eff}=4.7\times
10^{-10} \mathrm{s}^{-1}$, and
$A_\mathrm{eff}=1.9\times10^{-6} \mathrm{s}^{-1}$. 
A comparison of the heating rates using our effective
two-level system with the results using the 15-level molecule
in our KOSMA-$\tau$ model is shown in 
Figure~\ref{h2hn30}. Here, we scan the parameter ranges where
the H$_2$ vibrational heating and cooling gives a major contribution
to the overall energy balance (see Fig. \ref{bal_heat}). In the
upper plot representing low radiation fields and varying densities
we find an almost perfect agreement of the two level approximation
with the full numeric treatment. At high densities and varying
radiation fields shown in the lower  plot, we find a good match
at temperatures below about 3000~K. The deviation at higher
temperatures is due to the neglect of cooling contributions
from higher vibrational levels. It has no impact on the overall
PDR model, because photoelectric heating clearly supersedes the
vibrational contribution at these conditions.

With the simple analytic two-level approximation, we can
easily understand the quantitative behavior
of the H$_2$ vibrational energy balance from basic principles
thus providing a handy tool for estimates of temperature structures.
\citet{burton90} also introduced a
two-level approximation for the
H$_2$ heating. However, they considered only a single pseudo
excited level with an energy corresponding to $v=6$ and
hence did not properly account for cooling via
rapid excitation to $v=1$, or heating via pumping
and collisional deexcitation from all 15 levels.

\end{document}